# (Un)conscious Bias in the Astronomical Profession: Universal Recommendations to improve Fairness, Inclusiveness, and Representation

Alessandra Aloisi[1] and Neill Reid[1]

[1]*Space Telescope Science Institute, 3700 San Martin Drive, Baltimore, MD 21218, USA*

## Abstract

(Un)conscious bias affects every aspect of the astronomical profession, from scientific activities (e.g., invitations to join collaborations, proposal selections, grant allocations, publication review processes, and invitations to attend and speak at conferences) to activities more strictly related to career advancement (e.g., reference letters, fellowships, hiring, promotion, and tenure). For many, (un)conscious bias is still the main hurdle to achieving excellence, as the most diverse talents encounter bigger challenges and difficulties to reach the same milestones than their more privileged colleagues. Over the past few years, the Space Telescope Science Institute (STScI) has constructed tools to raise awareness of (un)conscious bias and has designed guidelines and goals to increase diversity representation and outcome in its scientific activities, including career-related matters and STScI sponsored fellowships, conferences, workshops, and colloquia. STScI has also addressed (un)conscious bias in the peer-review process by anonymizing submission and evaluation of Hubble Space Telescope (and soon to be James Webb Space Telescope) observing proposals. In this white paper we present a plan to standardize these methods with the expectation that these universal recommendations will truly increase diversity, inclusiveness and fairness in Astronomy if applied consistently throughout all the scientific activities of the Astronomical community.

## 1. Why Diversity is important in Science

Diversity matters both in the workplace and in everyday life! Diversity allows for a wider range of people, skills, talents, and perspectives to be involved. When this happens, fewer concepts are taken for granted, more questions are raised, and more viewpoints are taken into account in devising solutions. The key to a breakthrough is often the ability to see the problem from different perspectives, not simply "being smart" (Page, 2007). When groups of highly talented people work together to tackle a problem, what matters most in achieving a solution is their diversity, rather than their individual abilities (Ely & Thomas 2001; Parrotta, Pozzoli, & Pytlikova 2011; Ostergaard, Timmermans, & Kristinsson 2011). Diversity is not separate from increasing overall quality, but rather essential to achieve it.

Diversity is particularly important in science where knowledge progresses in teams and through breakthroughs (Temm 2008). "Diversity in science refers to cultivating talent and promoting the full inclusion of excellence across the whole social spectrum" (Gibbs 2014), in a way that includes people of, e.g., all genders, races, and backgrounds. To achieve excellence in science, we need to continually focus on hiring, cultivating, and retaining talent that would not be otherwise accessible (Carrell, Page, & West 2010).

## 2. Science and the Paradox of Meritocracy

Many view diversity efforts in science as contrary to the ideal of meritocracy, a system that rewards a combination of ability and effort in a way that according to the Merriam-Webster dictionary definition, "the talented are chosen and moved ahead on the basis of their achievement".

However, it is necessary to acknowledge the existence of the so-called "Paradox of Meritocracy". When a company emphasizes meritocratic values, those in managerial positions award, e.g., larger rewards (salary, promotions), to male employees than to equally performing female employees (Castilla 2008; Castilla & Bernard 2010). Those who believe they are the most objective (scientists) may actually exhibit the biggest bias as they do not feel it is necessary to monitor and scrutinize their own behavior.

### 2.1. Simulating Gender Bias Effects

The paradox of meritocracy and its interplay with gender bias has been modelled through computer simulations by Martell, Lane, & Emrich (1996). An organization starts with a 1:1 gender ratio at each level. All employees go through a performance evaluation cycle twice a year for 10 years. A 5% increase is introduced in the men's scores. A 15% turnover at all levels is considered with new positions filled by the best performers from the preceding level. As Fig.1 shows, after 10 years, the organization ends up unbalanced with only 30% and 25% women at the senior and executive levels (levels 7 and 8).

This is what happens to women at every stage of their scientific career in academia. Individuals at the intersection of multiple dimensions (e.g., women of color) will suffer from cumulative effects along all those dimensions.

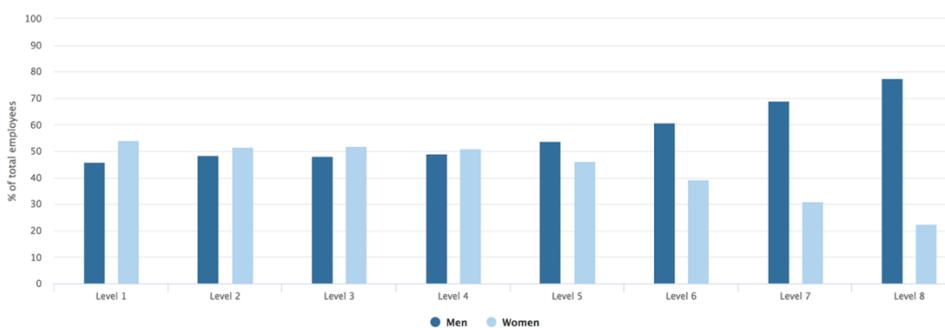

**Figure 1**: Changes in gender ratio for a 5% bias introduced in favor of men vs. women after 10 years in the simulation of an organization starting with a 1-1 ratio at each level (http://doesgenderbiasmatter.com/ which unfortunately is no longer available).

## 3. Deconstructing Bias

As scientists, we think we are objective. We have been trained to handle data objectively, and to apply the rigor of scientific method to everything we do. It is tempting to think that only a minority of uninformed people are biased. The truth is we are all biased. We all perceive and treat people based on their social groups.

We are continuously bombarded with a high volume of information and our brain relies on schemas to make sense of the world. Schemas are "cognitive shortcuts" used to increase efficiency in navigating situations, "conceptual frameworks" helping our brain to anticipate what to expect from experiences and situations. We are usually not aware that we have schemas, but our brain relies on them, particularly when we are under pressure and short in time. These "cognitive roadmaps" for quick processing and categorization of information, are built and enforced over a lifetime through personal experiences, social media and culture, and are widely shared.

Biases are schemas, (un)conscious hypotheses and expectations of others based on their group membership that influence our judgement and how we interact with them (regardless of our own group). These schemas include both "attitudes" - gut feelings towards a category (e.g., like vs. dislike) - and "stereotypes" - more specific associations between a category and a particular trait (e.g., Asians are good at math; women are natural caregivers). Attitudes/stereotypes introduce bias in the decision-making process, because they represent information that diverges from a neutral point. In particular, "explicit (conscious) biases" are attitudes/stereotypes that we are aware of and constantly check for accuracy, appropriateness, and fairness to self-correct as needed. "Implicit (unconscious) biases" are instead attitudes/stereotypes that we are not aware of, but are still influencing our actions and decisions. Ultimately, biases are errors in the decision-making. It is thus extremely important to be aware of them and mitigate their effects.

## 4. How to bring Unconscious Bias to light

"Project Implicit" is a non-profit organization focused on understanding bias. Their Implicit Association Test (IAT; https://implicit.harvard.edu/implicit/) is a web-based series of exercises that uses the reaction time for concepts associations to identify biases towards race or gender. E.g., the test may bring to surface that men are associated with science more than women.

## 5. Unconscious Bias Happens

Schemas do lead to unconscious bias and affect evaluation of individuals belonging to different groups by affecting the "standard" (Biernat & Manis 1994) and the "criteria" (Valian 1999) used to judge performance (Fig. 2).

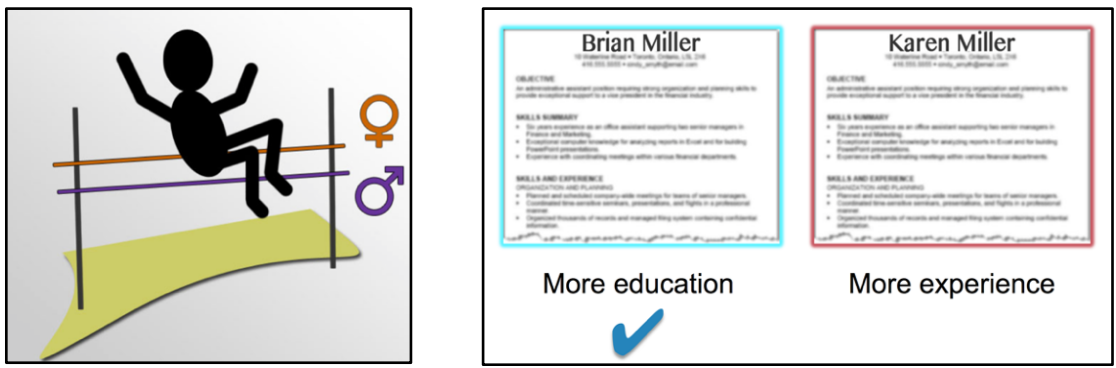

**Figure 2**: Two different ways in which unconscious bias affects evaluation of individuals (from "Fostering Women's Leadership", Davies' keynote presentation at "Building a Culture of Women Leaders" organized by the National District Attorney Association (NDAA) on 20 July 2015): women are held to higher "standards" than men *(left),* or different "criteria" are used to evaluate men and women *(right).*

In Astronomy, unconscious bias affects every aspect of the astronomical profession, from activities such as invitation to join collaborations, committee membership, peer-review process for articles, proposals and grants, and invitations to attend or speak at conferences, to activities related to career advancement, such as reference letters, applications for fellowships, hiring, promotion, and tenure.

Some examples of unconscious bias are given below. While not exhaustive, these examples contain lessons that we should learn and translate into new universal policies, if we want to make the field of Astronomy more inclusive and fairer.

### 5.1. Blind Music Auditions

The most famous example of unconscious bias is gender bias in orchestra auditions and hires. 'Blind" auditions (and later on a carpet) increased the probability for a woman to advance from preliminary rounds to subsequent rounds by 50%. It also accounted for 30% increase in the proportion of women hired in 1970-1996 (Golding & Rouse 2000).

### 5.2. Gender Bias in Evaluations of Curricula Vitae

By using the same CV assigned to a male or a female name, Steinpreis, Anders, & Ritzke (1999) found out that Brian Miller was hired two times more than Karen Miller. Moreover, reservations were expressed four times more often for Karen's tenure than for Brian's. This is consistent with earlier studies where female candidates were more likely hired as assistant professors and male candidates as associate professors (e.g., Fidell 1970). See Moss-Racusin et al. (2012) for a similar study on graduate student applications.

### 5.3. Race Bias in Evaluations of Resumes

By using the same resume assigned to a white-sounding name and African-American name, Bertrand & Mullainathan (2004) found out that Emily needed to send 10 resumes to get a callback, compared to Lakisha who needed to send 15 resumes, a significant 50% impact. On average, Lakisha needed 8 more years of experience to get as many callbacks as Emily. The higher the resume quality, the larger the gap between callbacks.

### 5.4. Gender Bias in Letters of Recommendation for Academia

Male applicants are more likely to get excellent letters and be described in terms of action-oriented characteristics (confident, assertive, aggressive, ambitious, independent). Terms like "brilliant scientist", "trailblazer", and "one of the best students I've ever had" are often used. Female applicants are more likely to get good letters and be described in terms of relationship-building characteristics (nurturing, caring, kind, agreeable, warm). Terms like "solid scientist doing good work", "highly intelligent", and "very knowledgeable" are used in this case. Women are usually not described in terms of the leadership skills necessary to excel in science, independently of writer's gender (Dutt et al. 2016; Madera, Hebl, & Martin 2009). This is penalizing when applying for jobs or coming up for tenure.

### 5.5. Biases in Peer-Review Process

Peer review is at the basis of science, yet (un)conscious bias heavily affects it. The introduction of dual-blind reviews resulted in an 8% increase (8% decrease) in first-author papers by female (male) authors (Budden et al. 2008). Similarly, when selecting contributions for conferences, there was a significant preference of single-blind reviewers to favor famous authors and authors from prestigious institutions compared to dual-blind reviewers (Tomkins, Zhang & Heavlin 2017).

## 6. Tackling (Un)conscious Bias at STScI

STScI houses ~ 100 research staff (equivalent to faculty in academia) who help support the space missions that the institute operates on behalf of NASA. STScI has several committees that manage a variety of scientific activities, from recruitment, evaluation, and promotion of research staff, to organization of the weekly colloquium and yearly scientific symposium. STScI has created tools to raise awareness of (un)conscious bias and designed guidelines and goals to increase diversity representation and outcome in its scientific activities.

### 6.1.   Diverse Committees for a Diverse Outcome

Based on the idea that a more diverse committee ensures a more diverse outcome (Casadevall & Handelsman 2014), STScI has adopted uniform guidelines on binary gender representation with a goal of 40% representation in each committee (similar to the fraction of women in early-career positions) and a minimum of 27% corresponding to the representation of women at STScI. We also created a baseline record based on the past 3-5 years that highlighted the bias present in our past activities, as well as metrics and tools to track progress towards our new goals. While the focus of the current work is on binary gender, we do believe that these best practices will significantly mitigate biases against other under-represented groups (De Rosa et al. 2019).

### 6.2.   Tailored (Un)conscious Bias Training

STScI has also compiled an extensive presentation on (un)conscious bias, with modules tailored to the specific activities of each scientific committee. This presentation is given at the start of each committee's activity.

### 6.3.   Newly Adopted Best Practices for Scientific Committees

Considerations from the ad-hoc presentation developed by STScI and recent practical experiences have converged into the formulation of the following recommendations as an effective tool to combat (un)conscious bias in the scientific committee activities at STScI: 1. increase bias awareness by recommending to take the IAT test; 2. define explicit and objective evaluation criteria prior to beginning deliberation; 3. read carefully all the materials submitted for each case; 4. be mindful of biases in recommendation letters; 5. give sufficient discussion time to each case; 6. be respectful of each other opinions; 7. use objective criteria and provide specific evidence in support of judgement; 8. refrain from identifying individuals by name and/or gender; rather use neutral terms like "the applicant", "the candidate", or "they" for individuals, and "the proposers" or "the proposal team" for proposals; 9. refrain from discussing PI's individual qualifications when discussing proposals; focus instead on the scientific and/or technical merits of the project.

For hiring, the following recommendations should also be considered: 1. increase number of women and minorities in the pool through active recruitment; 2. review applicant pool demographics with HR at every step of the selection process (triage, phone interviews, in-person interviews, offers); 3. use structured interviews - same questions and format for all candidates.

Even with a scientific committee that is diverse, has been trained about (un)conscious bias, and has followed the steps outlined above, the outcome may still be biased. Ex-officio members should be present at committee deliberations to ensure that best practices are followed. If the end result is still affected by bias, the recommendation is to raise awareness, discuss the biased outcome and redo the process.

### 6.4. Dual-Anonymous Review Process for HST Proposal Selection

STScI manages the activities related to the allocation of telescope time for HST and in the near future JWST. In the past, the HST proposal selection has shown evidence of gender bias with a success rate for women lower than for men (19% vs 23%; Reid 2014).

In order to counteract the gender and other biases (race, career stage, institutional size, and geographic origin), STScI introduced in 2018 a dual-anonymous review process. This first-of-its-kind process for proposals selection, resulted in women PIs having a higher success rate compared to men (8.7% versus 8.0%, respectively) for the very first time in 18 years (Fig. 3). This result was so encouraging that NASA is now implementing dual-anonymous reviews for all its facilities (Witze 2019).

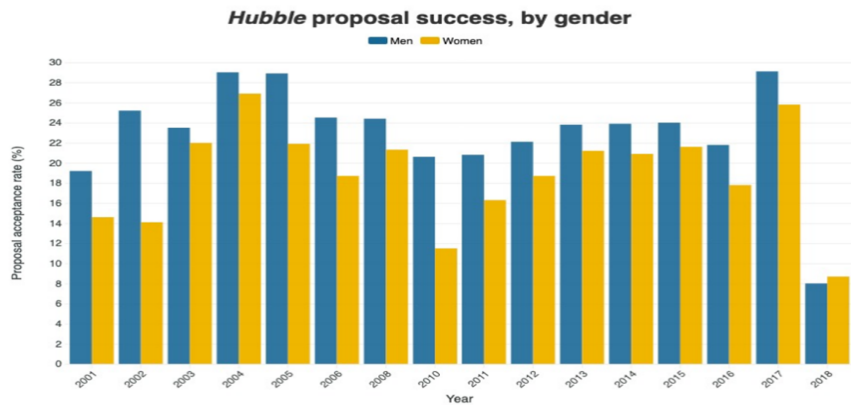

**Figure 3**: HST proposal acceptance rate by gender. From year 2001 to year 2017, there is a systematically higher success rate for men PIs than female PIs. This gender success rate drastically changed in 2018, when dual-anonymous peer-review process was introduced for the very first time in proposal selections to combat (un)conscious bias (Strolger & Natarajan 2019).

### 7. Recommendations and Conclusions

STScI has invested time and effort in the last ten years to mitigate the effect of implicit bias across its activities, to level the playing field for ALL its constituencies. This includes scientific activities (e.g., committees to generate speaker lists, colloquium series, Symposia), career advancement related activities (e.g., recruitment and personnel committees, hiring, promotion, and tenure), and peer-review activities (time allocation, grant allocation, fellowships).

Based on documented evidence in the literature and past and on-going STScI efforts to implement best practices to counteract (un)conscious bias, and encouraged by the positive results that these measures are achieving, we recommend the universal adoption of the following practices that will benefit all underrepresented groups in Astronomy.

- Introduce **bias training** as a necessary preamble to all activities that involve a committee where discussion is necessary and selective decisions are expected. The

- STScI experience is that bias training greatly improves bias awareness, and makes the subsequent discussion fairer and less prone to gross bias effects.
- Introduce **explicit and objective criteria** before starting the evaluation process and consistently use them throughout. Offer specific evidence in support of judgement. Advertisements for jobs should explicitly state the criteria used to select candidates.
- **Track progress**. A significant step in the process is to establish an institutional baseline of representation on committees in relation to their outcomes. This will allow each institution to verify whether the introduced measures are working and will document progress toward reaching the overall goal of a leveled and fair playing field.
- Introduce **dual-anonymous peer-review process**, not only for allocating **telescope time**, but also when disbursing **grants** and **funds**. Preliminary findings from the HST proposal selection processes in 2018-2020 show encouraging signs in several areas. As added bonus, reviewers remarked how the dual-anonymous process allowed them to focus on science and not the scientist, in what they called a "liberating" process.
- Introduce **dual-anonymous processes in recruitment**. While this idea is in its infancy, we recommend that institutions consider implementing pilot projects in a concerted fashion (maybe coordinated by the AAS). This will significantly improve the way short lists are made and candidates are recruited, and will greatly benefit minorities. This will also allow astronomers to review and revise (and maybe delete altogether) the process of submitting (biased) reference letters. We advocate exploring possibilities at institutional level and disseminating results nationwide.

Our expectation is that these universal recommendations will truly increase diversity, inclusion and fairness in Astronomy when applied consistently across all the scientific activities of the Astronomical community. We also advocate sharing and communicating experiences and results from the implementation of these practices widely so that we can all learn from each other. This effort can succeed only if the approach is shared in a collective fashion by the entire Astronomical community.